\documentclass{iauc}
\usepackage{graphicx}

\title[Regimes of Stability and Removal Time in the Asteroid Belt]
{Regimes of Stability and Scaling Relations for the Removal Time
in the Asteroid Belt: A Simple Kinetic Model and Numerical Tests}

\author{Mihailo \v{C}ubrovi\'{c}}

\affiliation{Department of Astronomy -- Petnica Science Center, P.
O. B. 6, 14104 Valjevo, Serbia and Montenegro \\ email:
cygnus@EUnet.yu}

\pubyear{2004} \volume{197} \pagerange{1--8}
\date{?? and in revised form ??}
 \jname{Dynamics of Populations of Planetary
Systems} \editors{Z. Kne\v{z}evi\'{c} and A. Milani, eds.}
\begin{document}

\maketitle

\begin{abstract}
We report on our theoretical and numerical results concerning the
transport mechanisms in the asteroid belt. We first derive a
simple kinetic model of chaotic diffusion and show how it gives
rise to some simple correlations (but not laws) between the
removal time (the time for an asteroid to experience a qualitative
change of dynamical behavior and enter a wide chaotic zone) and
the Lyapunov time. The correlations are shown to arise in two
different regimes, characterized by exponential and power-law
scalings. We also show how is the so-called "stable chaos"
(exponential regime) related to anomalous diffusion. Finally, we
check our results numerically and discuss their possible
applications in analyzing the motion of particular asteroids.
\keywords{minor planets, asteroids; diffusion; celestial
mechanics; methods: analytical}

\end{abstract}

\firstsection

\section{Introduction}
Despite some important breakthroughs in research of transport
mechanisms in the Solar system in the past decade, we still lack a
general \textit{quantitative} theory of chaotic transport, which
is especially notable for the so-called stable chaotic bodies. In
this paper, we sketch a new kinetic approach, which, in our
opinion, has a perspective of providing us such a theory sometime
in the future.

A kinetic model of transport has already been proposed by
\cite{out}. Although it is an important step forward, this model
fails to include a number of important effects. We also wish to
emphasize the role of phase space topology in the transport
processes. This has only recently been understood in papers by
\cite[Tsiganis, Varvoglis \& Hadjidemetriou (2000, 2002a,
2002b)]{127res, stabjov, kirkwood}. Still, the exact role of
cantori and stability islands in various resonances remains
unclear. This is one of the issues we intend to explore in this
paper. We argue that, due to the inhomogenous nature of the phase
space, a separate kinetic equation for each transport mechanism
should be constructed; after that, one can combine them to obtain
the description of long-time evolution. This is the basic idea of
our approach, which leads to some interesting statistical
consequences, such as anomalous diffusion and approximate scaling
of removal times with Lyapunov times.

\section{The kinetic scheme}

In order to model the transport, we use the "building block
approach" we have recently developed for Hamiltonian kinetics
(\cite[\v{C}ubrovi\'{c} 2004]{model}). We use the Fractional
Kinetic Equation (FKE), a natural generalization of the diffusion
equation for self-similar and strongly inhomogenous media
(\cite[e. g. Zaslavsky 2002]{physrep}):
\begin{equation}\label{fke}
\frac{\partial^{\beta}f\left(I, t\right)}{\partial
t^{\beta}}=\frac{\partial^{\alpha}}{\partial \vert
I\vert^{\alpha}}\left[\mathcal{D}\left(I\right)f\left(I,
t\right)\right]
\end{equation}
Thus, the evolution of the distribution function $f(I, t)$ is
governed by the transport coefficient $\mathcal{D}$ (the
generalization of the diffusion coefficient) and by the
(non-integer, in general) order of the derivatives $\alpha$
($0<\alpha\leq 2$) and $\beta$ ($0<\beta\leq 1$). The quantity
$\mu\equiv 2\beta/\alpha$ is called the transport exponent (for
the second moment the following holds asymptotically: $\langle
\Delta I^2\rangle\propto t^{\mu}$). If $\mu\neq 1$, the transport
is called anomalous (in contrast to normal transport or normal
diffusion\footnote{From now on, we will refer to any transport in
the phase space (i. e. evolution of the momenta of the action $I$)
as to "diffusion"; for the "classical" diffusion, we shall use the
term "normal diffusion".}).

We shall now very briefly describe each of the four building
blocks; unfortunately, most equations are cumbersome and
complicated, so we limit ourselves in this paper (for the sake of
conciseness) to merely state the basic ideas and final results of
the method. We use the planar MMR Hamiltonians
$H_{2BR}=H_{2BR}^0+H_{2BR}^{\prime}$ and
$H_{3BR}=H_{3BR}^0+H_{3BR}^{\prime}$ for two- and three-body
resonances, taken from \cite{3out} and \cite{3res}, respectively.
Under $H^0$ we assume the action-only part of the Hamiltonian.
$H_{2BR}$ was modified to account for the purely secular terms;
also, both Hamiltonians were modified to include the proper
precessions of Jupiter and Saturn\footnote{All our computations,
analytical and numerical, are performed with the osculating
elements, in order to gain as much simplification as possible.
However, in order to avoid the non-diffusive oscillations of the
osculating elements, one should use the proper elements instead;
we plan to do this in the future.}:
\begin{equation}\label{h2}
H_{2BR}^{\prime}=\sum_{m=0, 1; s=0\ldots k_J-k}^{u_5,
u_6}c_{msu_5u_6}\cos\left[m\left(
k_J\lambda_J-k\lambda\right)+sp+\left(u_5 g_5+u_6
g_6\right)t+u_5\beta_J+u_6\beta_S\right]\end{equation}\
\begin{equation}\label{h3} H_{3BR}^{\prime}=\sum_{m=0, 1; s}^{u_5,
u_6}c_{msu_5u_6}\cos\left[m\left(k_J\lambda_J+k_S\lambda_S+k\lambda\right)+sp+\left(u_5
g_5+u_6 g_6\right)t+u_5\beta_J+u_6\beta_S\right]
\end{equation}
The notation is usual. In $H_{2BR}^{\prime}$, we include all the
existing harmonics; in $H_{3BR}^{\prime}$, we include only those
given in \cite{3res}. In what follows, we shall consider only the
diffusion in eccentricity, i. e. $P$ Delauney variable. Inclusion
of the inclination could be important but we postpone it for
further work.

We estimate the transport coefficient as:

\begin{equation}\label{diff}
\mathcal{D}=\frac{T_{lib}^{(\alpha-\beta)}}{2}\sum_{s} s^{\alpha}
P^{s\alpha}\left(\sum_{u_5, u_6}
c^{\prime}_{0su_5u_6}(\alpha)+jc^{\prime}_{1s
u_5u_6}(\alpha)\right)
\end{equation}

where $T_{lib}$ denotes the libration period while
$c_{msu_5u_6}^\prime$ are coefficients dependent on the exponent
$\alpha$ from (\ref{fke}), independent on angles and $P$, which
were computed using the algorithm from \cite{dist}, for $H_{2BR}$,
or taken from \cite{3res}, for $H_{3BR}$. The indicator $j$ can be
equal to $0$ or $1$ (i. e. omission or inclusion of the resonant
terms), depending on the building block (see bellow). Although we
were able to compute also the higher-order corrections to this
quasilinear result in some cases, we neglect them in what follows,
in order to be able to solve the FKE analytically.

The first class of building blocks we consider are the overlapping
stochastic layers of subresonances. In this case, one expects a
free, quasi-random walk continuous in both time and space, since
no regular structures are preserved. Therefore, the FKE simplifies
to the usual diffusion equation, i. e. we have $\alpha=2$,
$\beta=1$ in (\ref{diff}); also, $j=1$ (the resonant harmonics are
actually the most important ones).

The above reasoning is only valid if the overlapping of
subresonances is not much smaller than $1$. Otherwise, the
diffusion can only be forced by the secular terms. Also, long
intervals between subsequent "jumps" induce the so-called "erratic
time", i. e. $\beta$ can be less than $1$, its value being
determined by the distribution of time intervals between "jumps"
$p(\Delta t)=1/\Delta t^{1+\beta}$. So, the transport coefficient
(\ref{diff}) now has $j=0$, $\alpha=2$ and $\beta<1$.

Our third class of building blocks are the resonant stability
islands. To estimate $\beta$ we use the same idea as in the
previous case; after that, we compute $\alpha$ from $\beta$ and
$\mu$, the transport exponent, which we deduce using the method
developed in \cite{rec}. Namely, analytical and numerical studies
strongly suggest a self-similar structure characterized by a
power-law scaling of trapping times $\lambda_T$, island surfaces
$\lambda_S$ and number of islands $\lambda_N$ at each level. The
transport exponent is then equal to
$\lambda_N\lambda_S/\lambda_T$. For some resonances and for some
island chains, we computed the scaling exponents applying the
renormalization of the resonant Hamiltonian as explained in
\cite{physrep}; in the cases when we did not know how to do this,
we used the relation between the transport exponent and the
fractal dimension $d_T$ of the trajectory in the $(P,p)$ space
(the space spanned by the action $P$ and the conjugate angle $p$),
which is actually the dimension of the Poincare section of the
trajectory:

\begin{equation}
d_T=\frac{2\lambda_T}{\lambda_N\lambda_S}
\end{equation}

The last remaining class of blocks are cantori. Here, we assume
the scaling of gap area on subsequent levels with exponent
$\lambda_S$ and an analogous scaling in trapping probability with
exponent $\lambda_p$, which determines the transport exponent as
$2\ln\lambda_p/\ln\lambda_S$; see also the reasoning from
\cite{shev}. The scaling exponents were estimated analogously to
the previous case.

For each building block, we construct a kinetic equation and solve
it. We always put a reflecting barrier at zero eccentricity and an
absorbing barrier at the Jupiter-crossing eccentricity. The
solution in Fourier space $(q, t)$ can be written approximately
in the following general form:

\begin{equation}\label{sol}
f_i(q,t)=E_{\beta}(-\vert
q^{\alpha}\vert\star\hat{\mathcal{D}_i}t^{\beta})
\end{equation}

where $E_{\beta}$ stands for the Mittag-Leffler function and
$\hat{\mathcal{D}_i}$ denotes the Fourier transform of
$\mathcal{D}_i$. The index $i$ denotes a particular building
block. The key for obtaining the global picture is to perform a
convolution of the solutions for all the building blocks.
Furthermore, one must take into account that the object can start
in different blocks and also that, sometimes, different ordering
of the visited blocks is possible. Therefore, one has the
following sum over all possible variations of blocks (we call it
Equation of Global Evolution - EGE):

\begin{equation}
f(P,t)=\sum\left[p_1 f_1(P,t)\star p_2 f_2(P,t)\star\dots \star
p_i f_i(P,t)\star\dots\right]
\end{equation}

To calculate it, one has to know also the transition probabilities
$p_i$, which is not possible to achieve solely by the means of
analytic computations. That is why we turn again to semi-analytic
results.

\section{Removal times, Lyapunov times and $T_L$ -- $T_R$ correlations}

The first task is to determine the relevant building blocks and
transitional probabilities. We do that by considering the
overviews of various resonances as given in \cite{moons1, moons2,
kirk21}. For the resonances not included in these references, we
turn again to the inspection of Poincare surfaces of section,
integrating the resonant models (\ref{h2}) and (\ref{h3}). In this
case, the probabilities are estimated as the relative measures of
the corresponding trajectories on the surface of section.

The result of solving the EGE is again a Mittag-Leffler function:

\begin{equation}
f_{global}(q,t)=E_{\gamma}(-\vert q\vert^\delta t^\gamma)
\end{equation}

The asymptotic behavior of this function, described e. g. in
\cite{physrep}, has two different forms: the exponential one and
the power-law one, depending on the coefficients $\gamma$ and
$\delta$, which are determined by the probabilities $p_i$ and
transport coefficients and exponents of the building blocks. In
the small $\gamma$ limit, the behavior is exponential and the
second momentum scales with $\tau_{cross}$, where $\tau_{cross}$
is the timescale of crossing a single subresonance, which we
interpret as the Lyapunov time\footnote{One should bear in mind
that this is just an approximation; strictly speaking, Lyapunov
time is not equal, nor simply related to the subresonance crossing
time.}. When $\gamma$ becomes large and the role of stickiness
more or less negligible, one gets a power-law dependance on
$\tau_{cross}$, i. e. $T_L$. So, we have the expressions:

\begin{equation}
T_R\propto\exp\left(T_L^x\right)\Phi\left(\Lambda_0,\cos(\ln
P_0),Q_0\right)
\end{equation}

for the exponential or stable chaotic regime, and:

\begin{equation}
T_R\propto\left(T_L^y\right)\Phi\left(\Lambda_0,\cos(\ln
P_0),Q_0\right)
\end{equation}

for the power-law regime. The scalings are not exact because the
fluctuational terms $\Phi\left(\Lambda_0,P_0,Q_0\right)$ appear.
These terms are log-periodic in $P_0$ and can explain the
log-normal tails of the $T_R$ distribution, detected numerically
e. g. in \cite{127res}.

\section{Results for particular resonances}

We plan to do a systematic kinetic survey of all the relevant
resonances in the asteroid belt. Up to now, we have only
preliminary results for some resonances.

Table 1 sums up our results for all the resonances we have
explored. For each resonance, we give our analytically calculated
estimates for $T_L$ and $T_R$. We always give a range of values,
obtained for various initial conditions inside the resonance. If
the "mixing" of the phase space is very prominent, we sometimes
get a very wide range, which includes both normal and stable
chaotic orbits. One should note that the "errorbars" in the plot
are simply the intervals of computed values -- they do not
represent the numerical errors. We also indicate if the resonance
has a resonant periodic orbit, which is, according to
\cite{kirkwood}, the key property for producing the \textit{fast}
chaos\footnote{The existence of the periodic orbit for $13:6$ and
$18:7$ resonances has not been checked thus far; however, we think
this would be highly unlikely for such high-order resonances}.
Bulirsch-Stoer integrator with Jupiter and Saturn as perturbers
was used for the integrations.

We have also tried to deduce the age of the Veritas family, whose
most chaotic part lies inside the $5 -2 -2$ resonance. Our EGE
gives an approximate age about 9 Myr while, assuming a constant
diffusion coefficient (Kne\v{z}evi\'{c}, personal communication),
one gets about 8.3 Myr. The similarity is probably due to the
young age of the family: were it older, the effects of
non-linearity would prevail and our model would give an age
estimate which is substantially different from that obtained in a
linear approximation.

\begin{table}
  \begin{center}
  \caption{Analytical and numerical values of the Lyapunov time (in Kyr) and removal time (in Myr). Existence of the periodic orbit in the planar
  problem for the 2BR is also indicated; the data in this column were taken from \cite{kirkwood}.}
  \begin{tabular}{llllll}\hline
  Resonance  & $T_L$ (Kyr) &   $T_L^{Num}$ (Kyr) & $T_R$ (Myr) & $T_R^{Num} (Myr)$ & Per. orbit ?\\
       $2:1$   & 1.1--3.4 & 2.9--5.2 & 0.8--19.4 & 1.1--31.2 & Yes \\
       $3:2$   & 1.4--3.5 & 3.1--6.2 & 0.9--19.1 & 3.2--142.7 & Yes \\
       $3:1$   & 6.2--8.2 & 6.8--9.4 & 0.8--8.0 & 1.0--21.2 & Yes \\
       $5:3$   & 1.9--2.2 & 1.1--3.5 & 0.9--4.1 & 0.8--7.4 & Yes \\
       $5:2$   & 6.9--9.2 & 8.3--14.4 & 7.6--13.4 & 9.2--28.7 & Yes \\
       $7:4$   & 2.7--5.3 & 1.8--3.9 & 6.9--23.4 & 12.2--41.3 & Yes \\
       $8:5$   & 3.7--6.7 & 4.1--7.6 & 8.2--18.2 & 7.3--23.4 & No \\
       $7:3$   & 5.0--6.9 & 6.9--9.1 & 16.7--62.2 & 22.4--91.2 & No \\
       $9:5$   & 5.4--6.8 & 3.1--5.2 & 23.2--86.7 & 11.3--104.2 & No \\
       $11:7$   & 7.1--14.1 & 3.4--9.7 & 10.6--73.4 & 8.4--88.2 & Yes \\
       $11:6$   & 10.8--22.7 & 10.1--16.5 & 16.4--330.3 & 13.2--$\approx 500$ & No \\
       $12:7$   & 11.1--14.7 & 4.1--15.2 & 4.6--278.1 & 5.2--$\approx 1000$ & No \\
       $13:7$   & 65.1--84.6 & 43.3--76.1 & 23.2-- $\approx 1000$& 14.2-- $>1000$ & No \\
       $13:6$   & 55.1--77.6 & 14.6--24.5 & 41.3-- $\approx 1000$ & 21.1-- $>1000$ & No \\
       $18:7$   & $\approx 500$ & 300-600 & $\approx 20000$ & $>1000$ & No \\
       $5-2-2$   & 11.4--13.7 & 8.4---10.9 & $\approx 10000$ & $>1000$ & No \\
       $2+2-1$   & 90--160 & 130--240 & $\approx 20000$ & $>1000$ & No \\
       $6+1-3$   & 130--150 & 130--170 & $\approx 40000$ & $>1000$ & No \\\hline
  \end{tabular}
  \end{center}
\end{table}

Figure 1 gives the results from table 1 plotted along the
semimajor axis. It can be noted that the agreement is good within
an order of magnitude, with some exceptions. Actually, one can see
that the disagreement with the simulations is most significant
exactly in the resonances with a periodic orbit, which might
actually require a completely different treatment of transport.

\begin{figure}
\includegraphics[height=3in,width=4in]{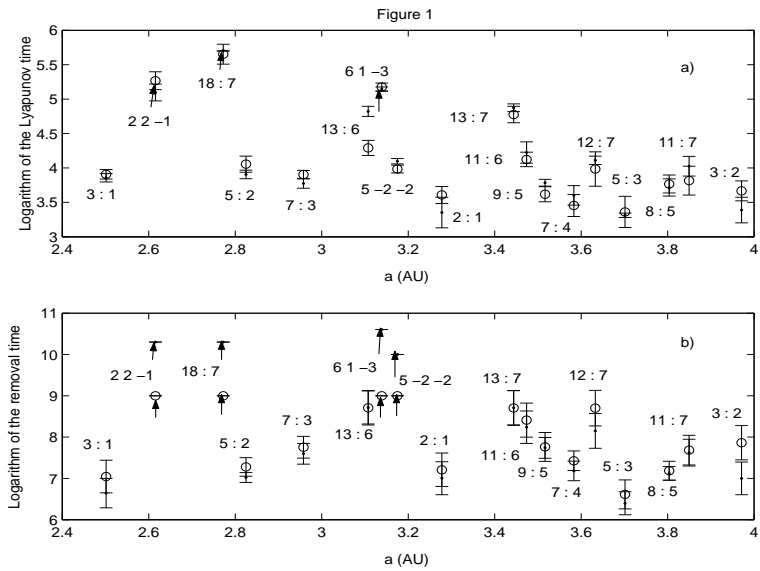}
  \caption{Analytical (points) and numerical (circles) values from table 1, for Lyapunov time (a) and removal time (b).
           Arrows correspond to extremely uncertain numerical values, usually the values close to or larger than the integration timespan (1 Gyr).}
\end{figure}

In figure 2, we plot the numerical $T_L$ -- $T_R$ relation for the
resonances $5:3$ and $12:7$, examples of normal and stable chaos,
respectively. The largest discrepancies in the figure 2a are
probably for the objects near the stability islands; in the figure
2b, the fit is completely ruined. To cheque the assumption that
this is due to the mixing of populations, we integrate a larger
population of objects and divide them into two classes (the
criterion being the prominence of anomalous diffusion, see later).
For each class, we perform a separate fit with the corresponding
$T_L$ -- $T_R$ relation. Now most objects can be classified into
one of the two scaling classes. In particular, this shows that the
famous stable-chaotic object 522 Helga is probably not a remnant
of some larger initial population but rather a member of one of
the two populations existing in this resonance.

\begin{figure}
\includegraphics[height=3in,width=4in]{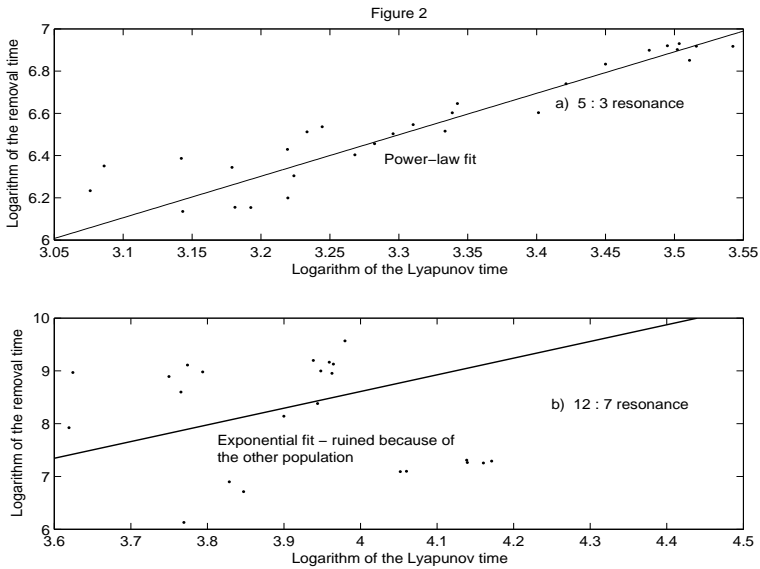}
  \caption{Plot of $T_L$ -- $T_R$ dependance for the resonances $5:3$ (a), fit with a power-law, and $12:7$ (b), fit with the exponential law.
           The exponential fit also looks linear, due to obviously bad fit quality. See text for comments.}
\end{figure}

\begin{figure}
\includegraphics[height=3in,width=4in]{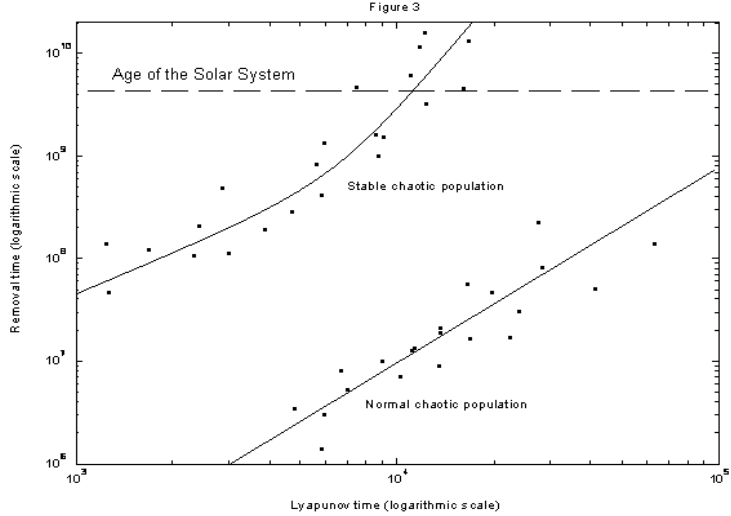}
  \caption{The same as in figure 2b but with two separate plots for the exponential regime population,
           and for the power-law regime population. Obviously, we have a substantially better fit.}
\end{figure}

Finally, in figure 4, we give the time evolution of the dispersion
in $P$ (i. e. $\langle\Delta P^2\rangle$) for a set of clones of
522 Helga, using the procedure described in \cite{diff}. Anomalous
diffusion is clearly visible. This confirms the stable chaotic
nature of this object and shows that we can use the anomalous
character of diffusion as an indicator of stable chaos.

\begin{figure}
\includegraphics[height=3in,width=4in]{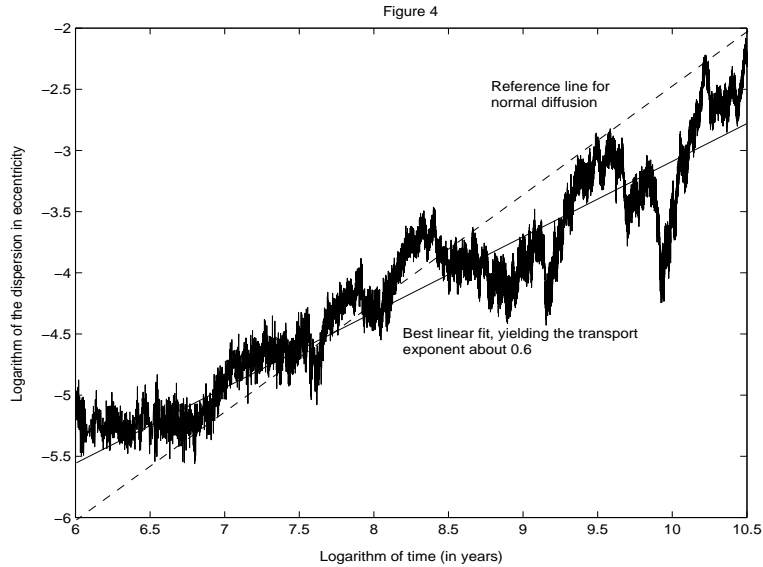}
  \caption{Diffusion in eccentricity for a set of clones of 522 Helga. Domination of anomalous transport is obvious.
           The reference line for normal diffusion ($\mu=1$) is also plotted.}
\end{figure}

\section{Conclusions and discussion}

We have given a kinetic model of chaotic transport in the asteroid
belt, based on the concept of convolution of various building
blocks. Combining numerical and analytical results, we have shown
how the removal time can be calculated and interrelated with the
Lyapunov time. We have obtained two regimes for chaotic bodies,
the power-law one and the exponential one. Due to the fractal
structure of the phase space, however, asteroids from different
regimes can be "mixed" in a small region of the phase space.

We would like to comment briefly on the controversial issue of the
$T_L$ -- $T_R$ relation. First of all, the correlations we have
found are \textit{of statistical nature only} and should not be
regarded as "laws" in the sense of \cite{frank94}. Furthermore,
due to their statistical nature, they cannot be used for any
particular object, only for populations. Finally, it is clear that
the scalings are \textit{non-universal}, i. e. the scaling
exponents are different for different resonances (possibly also in
disconnected regions of a single resonance).

The exponential regime, characterized mainly by anomalous
transport through various quasi-stable structures, corresponds to
the stable chaotic regime, discussed e. g. in \cite{127res} and
\cite{stabjov}. The reason that the exponential $T_L$ -- $T_R$
correlation was not noticed thus far are in part very large values
of $T_R$ in this regime, and in part the fact that stable chaotic
objects are typically mixed with the objects in the normal chaotic
regime. Also, it is interesting to note that the exponential
scalings are of the same form as those predicted in \cite{tetl}
for the Nekhoroshev regime; therefore, it seems that the
exponential stability can arise also due to stickyness, not
necessarily as a consequence of the Nekhoroshev structure.

Finally, we hope that our research will stimulate further work in
this field, since the results presented here are no more than just
a sketch of possible general theory.

\begin{acknowledgments}
I am greatly indebted to Zoran Kne\v{z}evi\'{c} for helpful
discussions and for permission to cite his yet unpublished
results. I am also grateful to George M. Zaslavsky, Harry
Varvoglis and Alessandro Morbidelli for sending me copies of some
of the references.
\end{acknowledgments}

\end{document}